# Thermal conductivity of benzothieno-benzothiophene derivatives at the nanoscale

*Magatte N. Gueye,[a,#] Alexandre Vercouter,[b,#] Rémy Jouclas,[c] David Guérin,[a] Vincent Lemaur,[b] Guillaume Schweicher[c,]*,[c] Stéphane Lenfant,[a] Aleandro Antidormi,[d] Yves Geerts,[c,e] Claudio Melis,[f] Jérôme Cornil[b,]* and Dominique Vuillaume[a,]**

a) Institute for Electronics Microelectronics and Nanotechnology (IEMN), CNRS, Av. Poincaré, Villeneuve d'Ascq, France.

b) Laboratory for Chemistry of Novel Materials, University of Mons, Place du Parc 20, Mons, Belgium.

c) Laboratoire de Chimie des Polymères, Faculté des Sciences, Université Libre de Bruxelles (ULB), Boulevard du Triomphe, 1050, Brussels, Belgium.

d) Catalan Institute of Nanoscience and Nanotechnology (ICN2), CSIC and BIST, Campus UAB, Bellaterra, 08193, Barcelona, Spain.

e) International Solvay Institutes for Physics and Chemistry, Brussels, Belgium.

f) Dipartimento di Fisica, Universita di Cagliari, Cittadella Universitaria, 09042 Monserrato (Ca), Italy.

We study by scanning thermal microscopy the nanoscale thermal conductance of films (40-400 nm thick) of [1]benzothieno[3,2-b][1]benzothiophene (BTBT) and 2,7-dioctyl[1]benzothieno[3,2-b][1]benzothiophene (C8-BTBT-C8). We demonstrate that the out-of-plane thermal conductivity is significant along the interlayer direction, larger for BTBT (0.63 ± 0.12 W m$^{-1}$ K$^{-1}$) compared to C8-BTBT-C8 (0.25 ± 0.13 W m$^{-1}$ K$^{-1}$). These results are supported by molecular dynamics calculations (Approach to Equilibrium Molecular Dynamics method) performed on the corresponding molecular crystals. The calculations point to significant thermal conductivity (3D-like) values along the 3 crystalline directions, with anisotropy factors between the crystalline directions below 1.8 for BTBT and below 2.8 for C8-BTBT-C8, in deep contrast with the charge transport properties featuring a two-dimensional character for these materials. In agreement with the experiments, the calculations yield larger values in BTBT compared to C8-BTBT-C8 (0.6-1.3 W m$^{-1}$ K$^{-1}$ versus 0.3-0.7 W m$^{-1}$ K$^{-1}$, respectively). The weak thickness dependence of the nanoscale thermal resistance is in agreement with a simple analytical model.

# INTRODUCTION.

Organic materials have recently attracted a great interest for potential thermoelectric application because commercial modules are built with materials such as bismuth telluride ($Bi_2Te_3$)-based alloys[1] which are toxic, expensive, and energy-consuming for processing. The candidates range from polymers (like PEDOT:PSS, PEDOT:OTf, and other derivatives) with high electrical conductivity and Seebeck coefficient (up to ~5,000-6,000 S cm$^{-1}$ and ~1 mV K$^{-1}$),[2-5] down to self-assembled monolayers and single molecule junctions (based mainly on alkyl chains, oligo(phenylene ethynylene)s, benzene, $C_{60}$,...).[6-13] In the latter case, quantum interference effects can be exploited to tailor and optimize the thermoelectric properties of the molecules.[14-17] In contrast to the charge transport properties, the thermal conductivity ($\mathcal{K}$) of thin films of small π-conjugated organic semiconductors (OSCs) appears less studied at both the experimental and theoretical levels. The in-plane $\mathcal{K}$ of various molecular thin films (pentacene, TPD, Alq3, $C_{60}$, PCBM, rubrene, DNTT, ....)[4, 18-26] has been measured in the range 0.1 - 0.8 W m$^{-1}$ K$^{-1}$ at the macroscale (ac-calorimetry, 3ω Joule heating, time domain thermo-reflectance....). Only a very few reports have been published on the out-of-plane $\mathcal{K}$ of these organic materials at the nanoscale, *e.g.* using the scanning thermal microscope (SThM).[27, 28] On the theoretical side, the prediction of $\mathcal{K}$ for OSCs and the definition of structure-property relationships is also scarcely addressed. The $\mathcal{K}$ values can be estimated by models[29] based on: (i) collective excitations of phonons via the resolution of the Boltzmann Transport Equation (BTE); (ii) atomic displacements via Molecular Dynamics (MD) approaches such as the Green-Kubo formalism or the Non Equilibrium Molecular Dynamics method (see below). The BTE approach is known to be far much prohibitive in terms of computational cost than MD-based techniques. Thus, $\mathcal{K}$ of a few molecules (pentacene, $C_{60}$, PCBM, $H_2Pc$, TPD,..)[30-38]

have been essentially calculated by various methods belonging to the "MD-class" in order to estimate the anisotropy of $\mathcal{K}$ along the long axis of the molecules versus in the perpendicular intralayer directions.

In this work, we measure by SThM the out-of-plane $\mathcal{K}$ at the nanoscale of thin films (40-400 nm thick) of [1]benzothieno[3,2-*b*][1]benzothiophene (BTBT) and 2,7-dioctyl[1]benzothieno[3,2-*b*][1]benzothiophene (C8-BTBT-C8).[39, 40] These molecules have promising performances for organic electronics with reproducible transistor hole mobilities in excess of 10 cm$^2$ V$^{-1}$ s$^{-1}$ and up to ~ 200 cm$^2$ V$^{-1}$ s$^{-1}$ at the local scale.[40, 41] Since their thermoelectric properties were not investigated experimentally up to now, this has stimulated the present experimental and theoretical works to determine the structure-property relationships of BTBT derivatives, especially the role of alkyl chains on the thermal transport. We demonstrate (SThM) that $\mathcal{K}_{BTBT}$ = 0.63 ± 0.12 W m$^{-1}$ K$^{-1}$ is larger than $\mathcal{K}_{C8-BTBT-C8}$ = 0.25 ± 0.13 W m$^{-1}$ K$^{-1}$. The nanoscale thermal resistance is weakly dependent on the film thickness, as predicted by a simple analytical model of the constriction thermal resistance for a surface coated by a thin film. The experimental results are supported by the theoretical estimates of $\mathcal{K}$ obtained by the AEMD (Approach to the Equilibrium Molecular Dynamics) method.[42] We find $\mathcal{K}$ along the long c-axis of the molecules larger for BTBT (1.04 W m$^{-1}$ K$^{-1}$) than for C8-BTBT-C8 (0.72 W m$^{-1}$ K$^{-1}$). The results also point to a decrease of $\mathcal{K}$ in the a-b plane upon alkylation (from 0.6-1.3 W m$^{-1}$ K$^{-1}$ for BTBT to 0.26-0.33 W m$^{-1}$ K$^{-1}$ for C8-BTBT-C8).

## RESULTS.

### *Scanning thermal microscopy results.*

Figure 1 shows the typical topographic and thermal voltage images of a C8-BTBT-C8 film prepared by spin-coating (see Methods). The organic thin film has a staircase topography (Fig. 1-a, height profile Fig. 1-c, red curve) with an incomplete surface coverage leaving apparent several zones of the underlying Si/SiO$_2$ substrate. The same type of "staircase" topography is observed for all samples (Figs. S1 and S2 in the ESI), in agreement with previous results.[43] In contrast, the SThM thermal voltage (Fig. 1-b) shows a featureless structure (Fig. 1-c, blue curve) for all samples (Figs. S1 and S2 in the ESI). The thermal voltage $V_{SThM}$ is related to the thermal resistance of the sample by:

$$V_{SThM} \propto \left(T_{sample} - T_{amb}\right) \propto R_{th} \dot{Q} \tag{1}$$

where $T_{sample}$ is the temperature of the surface sample, $T_{amb}$ the ambient temperature, $R_{th}$ the thermal resistance of the sample and $\dot{Q}$ the thermal flux. We analyzed the SThM image to determine the thermal resistance of the films at various locations with various thicknesses $t$, taking the thermal resistance of the Si/SiO$_2$ substrate as a reference (Fig. 1-d).

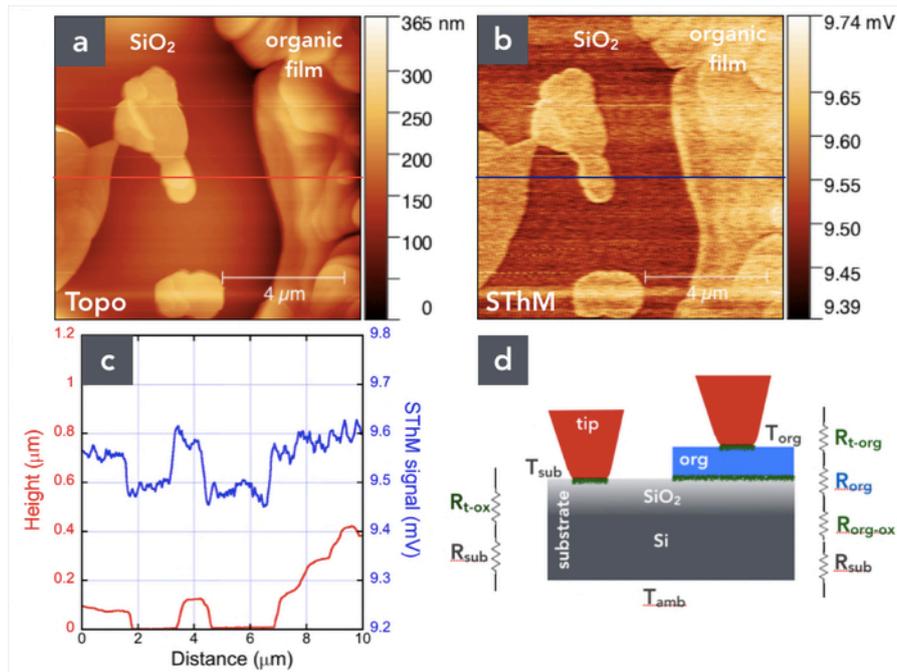

*Figure 1. (a) Topographic (10 µm x 10 µm) and (b) SThM thermal voltage (at the output of the Wheatstone bridge, 3ω-SThM method) images of a C8-BTBT-C8 film*

*spin-coated on Si/SiO$_2$ substrate, **(c)** height (red) and thermal voltage profiles (blue). **(d)** Schematic of the nanoscale SThM measurement. On the substrate, the thermal resistance is R$_{th-ox}$=R$_{t-ox}$+R$_{sub}$, with R$_{t-ox}$ the thermal resistance of the tip-SiO$_2$ interface and R$_{sub}$ the thermal resistance of the substrate. Over the organic domains, we have R$_{th-org}$=R$_{t-org}$+R$_{org}$(t)+R$_{org-ox}$+R$_{sub}$, with R$_{t-org}$ the thermal resistance of the tip-organic interface, R$_{org}$(t) the thermal resistance of the organic film of thickness t and R$_{org-ox}$ the thermal resistance of the organic-oxide interface (Fig. 1-d).*

We determined $\kappa$ using the null-point method, NP-SThM.[44] This differential method is suitable to remove the parasitic contributions (air conduction, etc…). When the tip contacts the sample surface (C), both the sample and parasitic thermal contributions are involved, whereas, just before physical tip contact (non contact, NC), only the parasitic thermal contributions are involved. We measured the thermal voltage V$_{SThM}$-z traces at several places on the organic films and on the substrate zones. The tip temperature is determined from V$_{SThM}$ (see the ESI). Figure 2 show 25 typical tip temperature versus distance (z-trace) curves measured on the C8-BTBT-C8 domain and on the nearby apparent Si/SiO$_2$ substrate (circled bullets on the SThM images in the insets). When approaching the heated tip to the surface, the tip temperature decreases gradually because the heat transfer through the air gap is increased. At contact, we observe a sudden decrease from T$_{NC}$ to T$_C$, due to the additional heat flux through the tip-sample contact. Figure 2-c shows the T$_C$ versus T$_{NC}$-T$_C$ curves for the C8-BTBT-C8 and BTBT samples and on the apparent substrate zone, where T$_C$ and T$_{NC}$ are the measured temperature at contact and non-contact conditions.

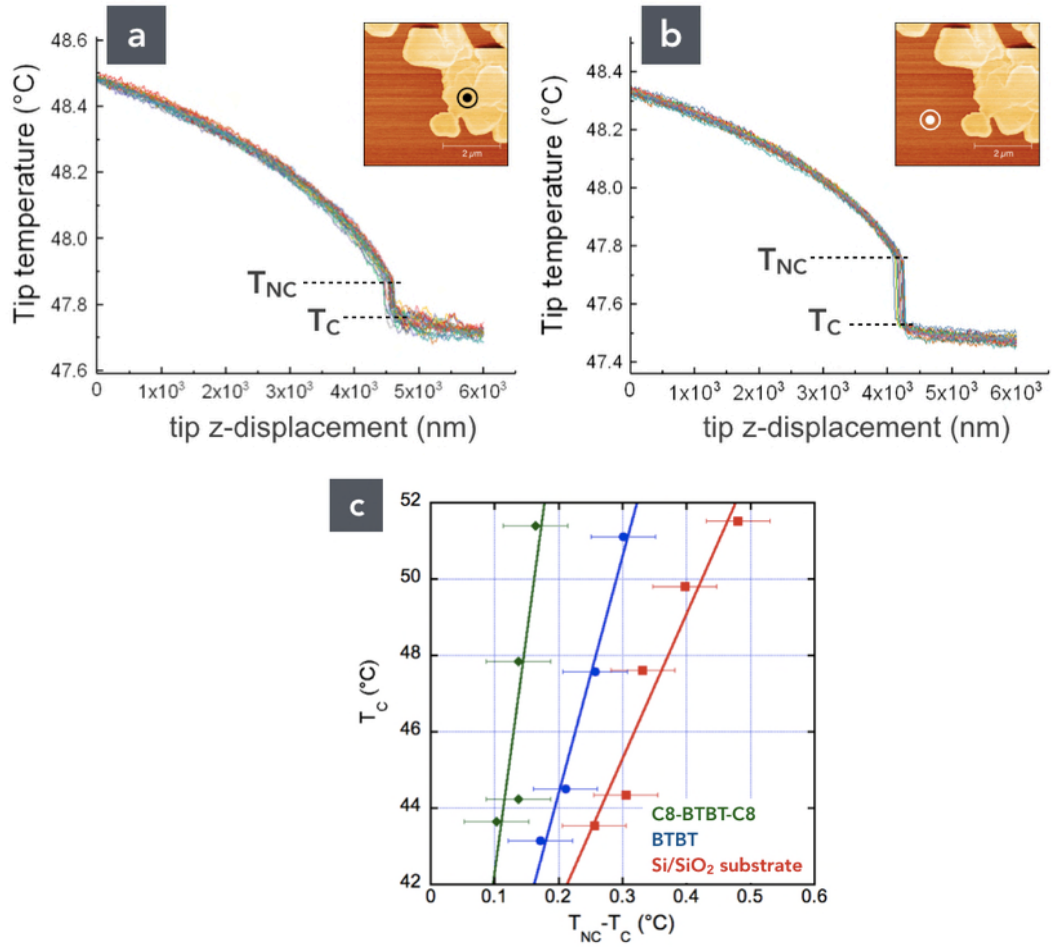

*Figure 2*. 25 tip temperature versus tip vertical displacement curves (z-trace, approach, 0 corresponds to the tip retracted) measured at $V_{DC}$ = 0.6 V on *(a)* a C8-BTBT-C8 domain (sample #5), *(b)* on the Si/SiO$_2$ substrate (as indicated by the circled bullets). *(c)* Temperature jump $T_{NC}$-$T_C$ versus temperature at tip contact $T_C$ for C8-BTBT-C8 (◆), BTBT (●) and the Si/SiO$_2$ substrate (■). The increasing $T_C$ corresponds to a supply voltage $V_{DC}$ from 0.6 - 0.9 V (by step of 0.1 V) of the Wheatstone bridge (0.6 - 1 V, step of 0.1 V on the substrate). Solid lines are the linear fits. Each data point (C8-BTBT-C8 and BTBT) is the average of 3 measurements at 3 different locations (25 $V_{SThM}$-z traces at each location). The data points for the substrate are averaged from the data acquired on the 2 samples.

The thermal conductivity is determined using the following relation:[44]

$$T_C - T_{amb} = \left[\alpha \frac{1}{\kappa} + \beta\right](T_{NC} - T_C) \qquad (2)$$

where α and β are calibrated parameters related to the SThM equipment and tip (Fig. S3 in the ESI) and $T_{amb}$ the room temperature (α = 25.6 W/m.K, β = 21.6 K/K and $T_{amb}$ = 22.5 °C). From a linear fit on these data, we get: $\kappa_{C8\text{-}BTBT\text{-}C8}$ = 0.25 ± 0.13 W m$^{-1}$ K$^{-1}$, $\kappa_{BTBT}$ = 0.63 ± 0.12 W m$^{-1}$ K$^{-1}$ and $\kappa_{sub}$ = 1.57 ± 0.43 W m$^{-1}$ K$^{-1}$. A key finding is that the thermal conductivity of BTBT is larger than that of C8-BTBT-C8, in full consistency with our AEMD simulations (see below). For the Si/SiO$_2$ substrate, we found a value close to that of bulk SiO$_2$, in agreement with previous measurements by NP-SThM[44] showing that the effective $\kappa$ is that of bulk SiO$_2$ (see Fig. 6-b in Ref. 44 and Fig. S4) if the SiO$_2$ thickness is larger than ~ 100 nm (here 200-500 nm).

The ratio of the thermal voltage measured on the organic domain over that on the substrate (Fig. 1), $V_{SThM\text{-}org}/V_{SThM\text{-}sub}$ is related to the ratio of the corresponding thermal resistance of each zone. For the Si/SiO$_2$ substrate, the constriction thermal resistance[45] is $R_{sub}=1/4r\kappa_{ox}$ = 9x10$^6$ K W$^{-1}$ with $\kappa_{ox}$ the "bulk" SiO$_2$ value (1.4 W m$^{-1}$ K$^{-1}$) and $r$ is the radius of the SThM tip thermal contact (estimated to be ≈ 20 nm, see the ESI). In order to determine the $R_{org}(t)$ from this SThM voltage ratio, we need to estimate the various interfacial resistances (Kapitza resistance)[46] which cannot be directly measured here. Reported values for a large variety of interfaces[27, 47-54] range typically between 10$^{-9}$ and 10$^{-6}$ m$^2$ K W$^{-1}$. For simplicity, we consider the lower limit value for all interfaces. In that case, the interface thermal resistances are negligible (8x10$^5$ K W$^{-1}$) compared to $R_{sub}$ and $R_{org}$ ($R_{sub}=1/4r\kappa_{ox}$ = 9x10$^6$ K W$^{-1}$, $R_{org}=1/4r\kappa_{org}$ =

2-5x10$^7$ K W$^{-1}$, considering the values of $\kappa_{org}$ determined above). In this oversimplified case, the ratio of the thermal voltage measured on the organic domain versus over the substrate is given by $V_{SThM-org}/V_{SThM-sub} = R^*_{org}(t)/R_{sub}$, assuming the same thermal flux $\dot{Q}$ on the oxide and the organic domain (see the ESI), where $R^*_{org}(t)$ is the effective constriction thermal resistance measured by the sharp SThM tip at the surface of the organic thin film. We consider a simple analytical model derived by Dryden[55] for film with $t/r>2$ (here $t > 40$ nm):

$$R^*_{org}(t) = \frac{1}{4r\kappa_{org}} - \frac{1}{2\pi r\kappa_{org}}\left(\frac{r}{t}\right)\ln\left(\frac{2}{1+\kappa_{org}/\kappa_{sub}}\right) \qquad (3)$$

The first term on the right-hand side stands for the constriction thermal resistance of the bulk organic film ($t \to \infty$) and the second term represents the effect of the underlying substrate covered by the film of thickness $t$. Figure 3 shows the measured $R^*_{org}(t)$ obtained from $V_{SThM-org}/V_{SThM-sub}$ ratios picked up at various locations on the organic films with various thicknesses (Fig. 1, Figs. S1 and S2, in the ESI) and on the underlying substrate, taking $R_{sub}=1/4r\kappa_{OX} = 9\times10^6$ K W$^{-1}$.

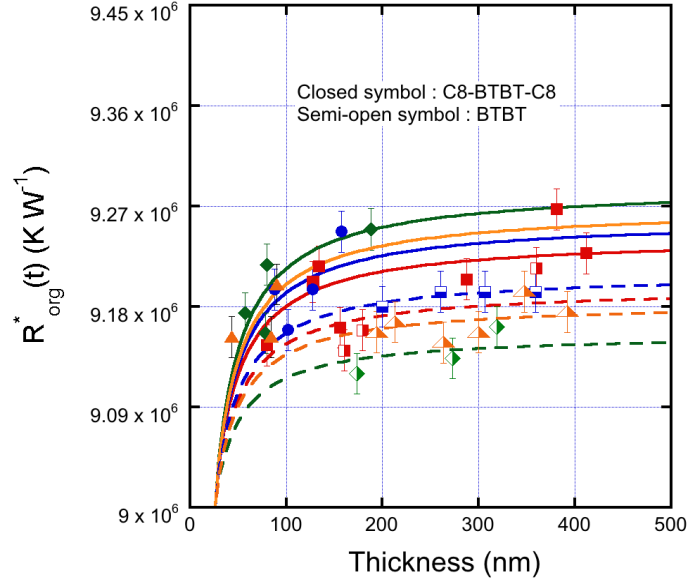

**Figure 3.** *Effective constriction thermal resistance $R^*_{org}(t)$ measured at the surface of the organic film for the C8-BTBT-C8 samples (closed symbols) and for the BTBT samples (semi-open symbols). Data and fit (lines): red = sample #1, blue = sample #2, green = sample #3, and orange = sample #4 as defined in the ESI.*

For $t \to \infty$, $R^*_{org}(t)$ tends to saturate at a larger value for C8-BTBT-C8 than for BTBT although the difference is weak. This trend is in agreement with the NP-SThM determination that $\kappa_{BTBT} > \kappa_{C8\text{-}BTBT\text{-}C8}$. From the fits of Eq. 3 (solid lines in Fig. 4), we obtain a mean value (see the ESI) $\kappa_{BTBT}$ = 1.37 ± 0.01 W m$^{-1}$ K$^{-1}$ and $\kappa_{C8\text{-}BTBT\text{-}C8}$ = 1.35 ± 0.01 W m$^{-1}$ K$^{-1}$ which are not significantly different. This implies that, even though the same trends are observed as with the NP-SThM method, this image analysis approach is perturbed by the interfacial thermal resistance, which more or less masks the actual values of the organic film thermal resistance. Previous works [27, 28, 49, 50, 54] also reported SThM tip-organic materials and organic-SiO$_2$ interfacial thermal resistances larger than 10$^7$ K W$^{-1}$ (or > ~ 10$^{-8}$ m$^2$ K W$^{-1}$), *i.e.*, larger than $R_{org}$ of our C8-BTBT-C8 and BTBT films.

***Theory.***

We use the Approach to the Equilibrium Molecular Dynamics (AEMD) method[42] to compute the lattice $\mathcal{K}$. We ignored the contribution of the electronic $\mathcal{K}$ at this stage because the thermal transport is barely controlled by electrons in most neutral and weakly doped organic semiconductors (OSCs).[56] This argument is supported by the newly developed molecular Wiedemann-Franz model[57] which predicts an electronic contribution to the heat transport smaller by several orders of magnitude than the corresponding lattice contribution. In brief, the key steps of the AEMD methodology are the following: (i) applying a perfectly monitored thermal pulse on a simulation box; (ii) recovering the initial thermodynamic equilibrium during a fast transient regime; (iii) fitting the time-decaying temperature difference between the right and left parts of the system from a reliable solution of the one-dimensional heat equation $\frac{\partial T}{\partial t}=D\frac{\partial^2 T}{\partial x^2}$ in order to evaluate the thermal diffusivity, $D$.[58] This alternative scheme has the benefit to be far less time-consuming than the previously cited methods due to the rapid dissipation of the thermal gradient. Then $\mathcal{K}=D\rho C_P$ is finally obtained, provided the density, $\rho$, and the specific heat, $C_P$, of the system are known (see details in the ESI). It is also of prime importance to account for the size-dependence of the $\mathcal{K}$ values deduced from this approach since phonons having a mean free path larger than the cell dimension do not effectively contribute to $\mathcal{K}$. An extrapolation procedure is thus needed to get rid of these size effects[59] and to extract a quasi-length-free lattice $\mathcal{K}$ from the linear regression of *1/$\mathcal{K}$* versus *1/L* (*L* the length of the box size along the direction of heat propagation).

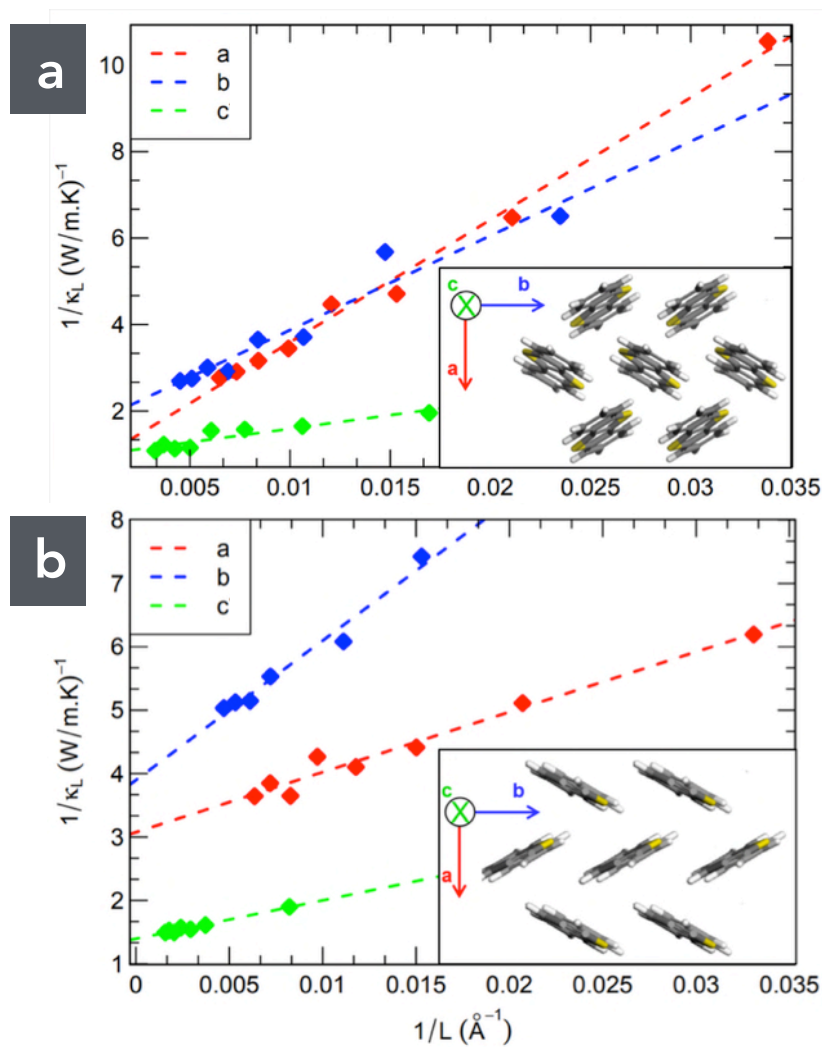

*Figure 4.* Inverse of the lattice thermal conductivity as a function of the inverse of the simulation box length along directions a, b and c for **(a)** BTBT and **(b)** C8-BTBT-C8. Each inset represents the molecular packing of BTBT and C8-BTBT-C8 in the ab plane. For sake of clarity, we did not represent the alkyl chains in (b). The two compounds crystallize into a monoclinic structure (a = 5.854 Å, b = 7.960 Å, c = 11.809 Å, α = 105.990° for BTBT and a = 5.927 Å, b = 7.880 Å, c = 29.180 Å, β = 92.443° for C8-BTBT-C8; respectively); both of them exhibit a very similar layered herringbone packing,[60] as shown in the insets.

Figure 4 display the inverse of the calculated lattice $\kappa$ as a function of the inverse of the super cell length. The linear extrapolation procedure provides $\kappa$ values of 1.31, 0.59 and 1.04 W m$^{-1}$ K$^{-1}$ along directions a, b and c for BTBT and 0.33, 0.26 and 0.72 W m$^{-1}$ K$^{-1}$ along the same directions for C8-BTBT-C8 (Table 1). There is a fairly good quantitative agreement between experimental measurements and theoretical estimates. Interestingly, the calculations indicate that the overall calculated conductivity (along the 3 crystal axes) of BTBT is higher than for C8-BTBT-C8 (Table 1), in full consistency with the experimental measurements probing essentially heat transport along the c axis.

| Theory | BTBT | C8-BTBT-C8 |
|---|---|---|
| a-axis | 1.31 | 0.33 |
| b-axis | 0.59 | 0.26 |
| c-axis | 1.04 | 0.72 |
| isotropic approximation | 0.95 | 0.46 |
| Experiment | | |
| null-point SThM | 0.63 ± 0.12 | 0.25 ± 0.13 |

**Table 1**. *Calculated and measured thermal conductivity values (W m$^{-1}$ K$^{-1}$)*

## DISCUSSION.

Since no theoretical and experimental data of the out-of-plane $\kappa$ was previously reported for these BTBT derivatives, we compare our values with other results obtained for various OSCs. The measured nanoscale out-of-plane $\kappa_{C8-BTBT-C8}$ = 0.25 ± 0.13 W m$^{-1}$ K$^{-1}$, is on par with values reported for other small molecule thin films by SThM: 0.15-0.20 W m$^{-1}$ K$^{-1}$ for methylstyryl-benzene,[28] 0.15-0.4 W m$^{-1}$ K$^{-1}$ for CuPc (Cu-phthalocyanine) and 0.15-0.25 W m$^{-1}$ K$^{-1}$ for PbCu.[27] They are also consistent with the corresponding values obtained by characterization methods at the macroscale on rubrene (0.07 W m$^{-1}$ K$^{-1}$)[23] and DNTT (0.45 W m$^{-1}$ K$^{-1}$).[24] The experimental out-of-plane $\kappa_{BTBT}$ = 0.63 ± 0.12 W m$^{-1}$ K$^{-1}$ lies at the highest limit of reported values: 0.45 W m$^{-1}$ K$^{-1}$ for DNTT[24] (to the best of our knowledge and excluding the highly unusual value of 21 W m$^{-1}$ K$^{-1}$ reported for crystal of TIPS-pentacene[61, 62]).

The calculations of the lattice $\kappa$ were performed on single crystals while the SThM measurement was carried out on polycrystalline samples containing a certain amount of disorder. It is also known (X-ray diffraction) that a disordered layer of less than *ca.* 10 nm exists at the SiO$_2$/BTBT interfaces.[39] To cope with this "experimental" disorder we compare the measured values with an effective isotropic $\kappa_{eff}$ defined as[63] $\kappa_{eff} = \sqrt{\kappa_c \kappa_{ab}}$ where the in-plane $\kappa$ is $\kappa_{ab} = \sqrt{\kappa_a \kappa_b}$. The calculated values are given in Table 1. A ratio of 2-2.5 between the thermal conductivities of BTBT and C8-BTBT-C8 is observed at both the experimental and theoretical level. For both materials, the agreement between the evaluated $\kappa_{eff}$ and the measurements (Table 1) is excellent, supporting the existence of a thin disordered layer at the SiO$_2$/BTBT interface with a lower thermal conductivity (*i.e.* ab-plane).

We do find a rather isotropic (3D) behavior of the heat propagation in the two derivatives in deep contrast with the charge transport properties featuring a two-dimensional character in presence of a herringbone arrangement;[60, 64, 65] note that a significant thermal conductivity along the three crystallines axes has also been reported theoretically for the DNTT single crystal[36]. The $\kappa$ ratios for a/b, c/a and c/b are respectively 2.22, 0.79 and 1.76 for BTBT and 1.27, 2.18 and 2.77 for C8-BTBT-C8. Morever, we evidence a noticeable drop in the ab-plane $\kappa$ when octyl chains are added on each side of the aromatic cores, with anisotropic factors of $\kappa_a^{C0}/\kappa_a^{C8}$ ~ 3.97 and $\kappa_b^{C0}/\kappa_b^{C8}$ ~ 2.27. This complements the theoretical study of Shi et al. [32] showing that $\kappa$ is marginally affected by the presence of terminal saturated chains in Cn-BTBT-Cn (with n = 8;10;12). It is worth stressing that the drop in thermal conductivity upon addition of saturated chains has been also clearly observed in recent $\kappa$ measurements performed on DNTT and C8-DNTT-C8 thin films in the a-b plane, and fully supported again by our AEMD calculations.[66]

To gain a deeper insight into the underlying physical mechanisms, we explore the spatial character of the vibrational modes, by estimating their participation ratio (PR).[67] This parameter is a quantitative measure of the spatial extension of vibrations, allowing to classify them into extended (large PR) and localized modes (PR $\sim$ 0). It has been shown[68] that extended modes are generally more effective in transporting heat across the material than localized modes. The details of the calculations are given in the ESI.

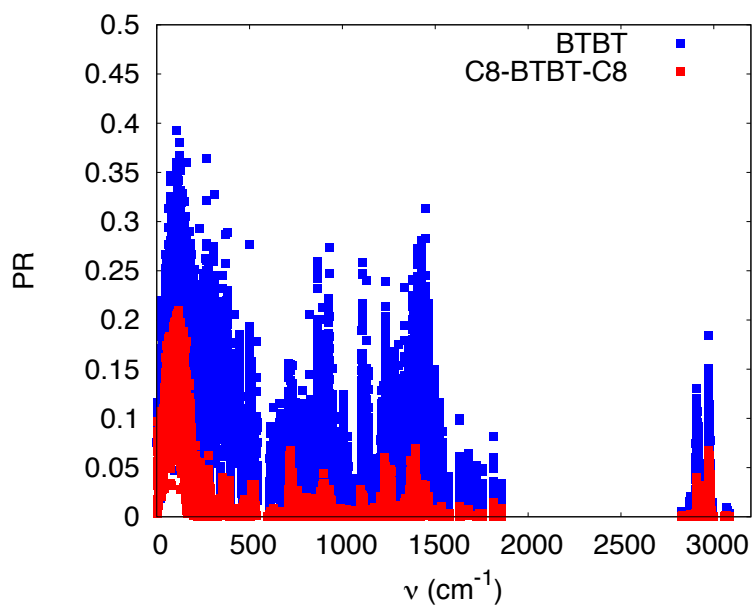

*Figure 5.* *Estimated Participation Ratio (PR) as a function of the vibrational frequency for BTBT (blue) and C8-BTBT-C8 (red).*

The participation ratio of vibrations in BTBT and C8-BTBT-C8 are shown in Fig. 5 as a function of their frequency. We observe that PR in BTBT takes larger values than C8-BTBT-C8 over the whole frequency spectrum. This shows that BTBT hosts vibrational modes are spatially more extended than those in C8-BTBT-C8. Hence, BTBT alkylation leads to a strong localization of the vibrational modes and consequently to a strong reduction of the overall thermal conductivity.

## CONCLUSION.

In summary, we demonstrate that the thermal conductivity of BTBT, $\kappa_{BTBT}$ = 0.63 ± 0.12 W m$^{-1}$ K$^{-1}$ is larger than for C8-BTBT-C8, $\kappa_{C8-BTBT-C8}$ = 0.25 ± 0.13 W m$^{-1}$ K$^{-1}$. The nanoscale thermal resistance is weakly dependent on the film thickness, as predicted by a simple analytical model of the constriction thermal resistance for a surface coated by a thin film. The experimental results are further supported by the theoretical estimates of the thermal conductivity obtained by the AEMD (Approach to the Equilibrium Molecular Dynamics) method. Moreover, we have not only demonstrated the drop of the thermal conductivity upon introduction of saturated chains but we have also provided a clear rationale based on the degree of delocalization of the intermolecular vibrational modes. Finally, our calculations point to significant thermal conductivity (3D-like) values along the 3 crystalline directions, with anisotropy factors between the crystalline directions below 1.8 for BTBT and below 2.8 for C8-BTBT-C8, in deep contrast with the charge transport properties featuring a two-dimensional character for these materials.

## METHODS.

***Synthesis and sample fabrication.*** BTBT and C8-BTBT-C8 were synthesized according to the reported procedures.[69, 70] The thin films were deposited by spin-coating on thermal $SiO_2$/Si-n+ according to previously reported methods.[39, 43] Nine samples (5 C8-BTBT-C8, 4 BTBT) were prepared using various spin-coating parameters to vary the film thickness (see the ESI).

*Scanning Thermal Microscope (SThM).* SThM[71, 72] were carried out with a Bruker ICON machine equipped with the Anasys SThM module. All measurements were done at room temperature in an air-conditioned laboratory ($T_{amb}$ = 22.5 °C, relative humidity of 35-40 %). In the scanning mode, the topography and thermal voltage were recorded simultaneously. We used both the DC method or the 3ω-SThM.[73] The thermal conductivity is determined by using the null-point SThM method (see the ESI).[44]

*Calculations of the lattice thermal conductivity.* All MD simulations have been performed in the LAMMPS software package[74] by means of the Optimized Potentials for Liquid Simulations All-Atoms (OPLS-AA) force-field (see the ESI). In addition, we also estimated the participation ratio to characterize the vibrational properties of both BTBT and C8-BTBT-C8. This parameter is a quantitative measure of the spatial extension of vibrations (see the ESI).

## ASSOCIATED CONTENT

Electronic Supplementary Information (ESI) is available:

Sample fabrication, SThM methods, theoretical methods, additional SThM images, NP-SThM calibration, estimation of the constriction thermal resistance of the Si/$SiO_2$ substrate, estimation of the radius of the thermal contact between tip and surface, fitted parameters of Eq. 3.

## AUTHOR CONTRIBUTIONS

M.N.G. prepared the samples, carried out and analyzed the SThM measurements with the help of D.G. and S.L., respectively. A.V. carried out the calculations of the thermal conductivity under the supervision of V.L. and J.C.. R.J., G.S. and Y.G. synthesized and assessed purity of the materials. A.A. and C.M. carried out the calculations of the vibrational properties. D.V. conducted the project, analyzed the results and wrote the paper with the help of all the authors. All authors have given approval to the final version of the manuscript.

# These authors (M.N.G. and A.V.) contributed equally to this work.

## CONFLICTS OF INTEREST.

The authors declare no competing financial interest.

## ACKNOWLEDGEMENTS.

M.G., D.G., S.L. and D.V. thank the ANR for financial support (ANR-16-CE05-0029). Y.G. is thankful to the Belgian National Fund for Scientific Research (FNRS) for financial support through research projects BTBT (# 2.4565.11), Phasetrans (# T.0058.14), Pi-Fast (# T.0072.18), 2D to 3D (# 30489208), and DIFFRA (# U.G001.19). Financial supports from the French Community of Belgian (ARC # 20061) is also acknowledged. G.S. acknowledges postdoctoral fellowship support from the FNRS. The work in the Laboratory for Chemistry of Novel Materials was supported by the Consortium des Équipements de Calcul Intensif (CÉCI), funded by the Fonds de la Recherche Scientifique de Belgique (F.R.S.-FNRS) under Grant # 2.5020.11. J.C. and A.V. are FNRS research fellows.

## REFERENCES.

1. H. J. Goldsmid, in *Semiconductors and Semimetals*, ed. T. M. Tritt, Elsevier, 2001, vol. 69, pp. 1-24.
2. O. Bubnova, Z. U. Khan, A. Malti, S. Braun, M. Fahlman, M. Berggren and X. Crispin, *Nat. Mater.*, 2011, **10**, 429-433.
3. M. N. Gueye, A. Carella, N. Massonnet, E. Yvenou, S. Brenet, J. Faure-Vincent, S. Pouget, F. Rieutord, H. Okuno, A. Benayad, R. Demadrille and J.-P. Simonato, *Chem. Mater*, 2016, **28**, 3462-3468.

4.  B. Russ, A. Glaudell, J. J. Urban, M. L. Chabinyc and R. A. Segalman, *Nature Reviews Materials*, 2016, **1**, 16050.
5.  M. N. Gueye, A. Carella, J. Faure-Vincent, R. Demadrille and J.-P. Simonato, *Progress in Materials Science*, 2020, **108**, 100616.
6.  M. Paulsson and S. Datta, *Phys. Rev. B*, 2003, **67**, 241403(R).
7.  P. Reddy, S.-Y. Jang, R. A. Segalman and A. Majumdar, *Science*, 2007, **315**, 1568-1571.
8.  J. A. Malen, P. Doak, K. Baheti, T. D. Tilley, R. A. Segalman and A. Majumdar, *Nano Lett*, 2009, **9**, 1164-1169.
9.  C. Evangeli, K. Gillemot, E. Leary, M. T. González, G. Rubio-Bollinger, C. J. Lambert and N. Agraït, *Nano Lett*, 2013, **13**, 2141-2145.
10. T. Meier, F. Menges, P. Nirmalraj, H. Hölscher, H. Riel and B. Gotsmann, *Phys. Rev. Lett.*, 2014, **113**, 060801.
11. L. Rincon-Garcia, C. Evangeli, G. Rubio-Bollinger and N. Agraït, *Chem. Soc. Rev.*, 2016, **45**, 4285-4306.
12. L. Cui, S. Hur, Z. A. Akbar, J. C. Klöckner, W. Jeong, F. Pauly, S.-Y. Jang, P. Reddy and E. Meyhofer, *Nature*, 2019, **572**, 628-633.
13. N. Mosso, H. Sadeghi, A. Gemma, S. Sangtarash, U. Drechsler, C. Lambert and B. Gotsmann, *Nano Lett*, 2019, **19**, 7614-7622.
14. C. J. Lambert, *Chem. Soc. Rev.*, 2015, **44**, 875-888.
15. J. C. Klöckner, J. C. Cuevas and F. Pauly, *Phys. Rev. B*, 2017, **96**, 245419.
16. R. Miao, H. Xu, M. Skripnik, L. Cui, K. Wang, K. G. L. Pedersen, M. Leijnse, F. Pauly, K. Wärnmark, E. Meyhofer, P. Reddy and H. Linke, *Nano Lett*, 2018, **18**, 5666-5672.
17. P. Gehring, J. M. Thijssen and H. S. J. Van Der Zant, *Nature Reviews Physics*, 2019, **1**, 381-396.
18. R. C. Yu, N. Tea, M. B. Salamon, D. Lorents and R. Malhotra, *Phys. Rev. Lett.*, 1992, **68**, 2050-2053.

19. N. Kim, B. Domercq, S. Yoo, A. Christensen, B. Kippelen and S. Graham, *Appl. Phys. Lett.*, 2005, **87**, 241908.
20. Y. Okada, M. Uno, Y. Nakazawa, K. Sasai, K. Matsukawa, M. Yoshimura, Y. Kitaoka, Y. Mori and J. Takeya, *Phys. Rev. B*, 2011, **83**, 113305.
21. J. C. Duda, P. E. Hopkins, Y. Shen and M. C. Gupta, *Phys. Rev. Lett.*, 2013, **110**, 015902.
22. X. Wang, C. D. Liman, N. D. Treat, M. L. Chabinyc and D. G. Cahill, *Phys. Rev. B*, 2013, **88**, 075310.
23. H. Zhang and J. W. Brill, *J. Appl. Phys.*, 2013, **114**, 043508.
24. X. Wang, K. D. Parrish, J. A. Malen and P. K. L. Chan, *Scientific Reports*, 2015, **5**, 16095.
25. Y. Yao, M. Shahi, M. M. Payne, J. E. Anthony and J. W. Brill, *J. Mater. Chem. C*, 2016, **4**, 8817-8821.
26. T. Nomoto, S. Imajo, S. Yamashita, H. Akutsu, Y. Nakazawa and A. I. Krivchikov, *Journal of Thermal Analysis and Calorimetry*, 2018, **135**, 2831-2836.
27. D. Trefon-Radziejewska, J. Juszczyk, A. Fleming, N. Horny, J. S. Antoniow, M. Chirtoc, A. Kaźmierczak-Bałata and J. Bodzenta, *Synthetic Metals*, 2017, **232**, 72-78.
28. Y. Zhang, C. Zhang, D. Wei, X. Bai and X. Xu, *CrystEngComm*, 2019, **21**, 5402-5409.
29. G. Fugallo and L. Colombo, *Physica Scripta*, 2018, **93**, 043002.
30. D. Wang, L. Tang, M. Long and Z. Shuai, *J. Phys. Chem. C*, 2011, **115**, 5940-5946.
31. J. Chen, D. Wang and Z. Shuai, Journal of Chemical Theory and Computation, 2012, **8**, 3338-3347.
32. W. Shi, J. Chen, J. Xi, D. Wang and Z. Shuai, *Chem. Mater*, 2014, **26**, 2669-2677.
33. L. Chen, X. Wang and S. Kumar, *Scientific Reports*, 2015, **5**, 12763.
34. J. Y. Kim and J. C. Grossman, *Nano Lett*, 2016, **16**, 4203-4209.

35. A. Giri and P. E. Hopkins, *J. Phys. Chem. Lett.*, 2017, **8**, 2153-2157.
36. X. Wang, J. Zhang, Y. Chen and P. K. L. Chan, *Nanoscale*, 2017, **9**, 2262-2271.
37. H. Kojima, M. Nakagawa, R. Abe, F. Fujiwara, Y. Yakiyama, H. Sakurai and M. Nakamura, *Chemistry Letters*, 2018, **47**, 524-527.
38. R. Sasaki, Y. Takahashi, Y. Hayashi and S. Kawauchi, *J. Phys. Chem. B*, 2020, **124**, 881-889.
39. G. Gbabode, M. Dohr, C. Niebel, J.-Y. Balandier, C. Ruzié, P. Négrier, D. Mondieig, Y. H. Geerts, R. Resel and M. Sferrazza, *ACS Appl. Mater. Interfaces*, 2014, **6**, 13413-13421.
40. Y. Tsutsui, G. Schweicher, B. Chattopadhyay, T. Sakurai, J.-B. Arlin, C. Ruzié, A. Aliev, A. Ciesielski, S. Colella, A. R. Kennedy, V. Lemaur, Y. Olivier, R. Hadji, L. Sanguinet, F. Castet, S. Osella, D. Dudenko, D. Beljonne, J. Cornil, P. Samorì, S. Seki and Y. H. Geerts, *Adv. Mater.*, 2016, **28**, 7106-7114.
41. G. Schweicher, V. Lemaur, C. Niebel, C. Ruzié, Y. Diao, O. Goto, W.-Y. Lee, Y. Kim, J.-B. Arlin, J. Karpinska, A. R. Kennedy, S. R. Parkin, Y. Olivier, S. C. B. Mannsfeld, J. Cornil, Y. H. Geerts and Z. Bao, *Adv. Mater.*, 2015, **27**, 3066-3072.
42. C. Melis, R. Dettori, S. Vandermeulen and L. Colombo, *The European Physical Journal B*, 2014, **87**, 96.
43. M. Dohr, H. M. A. Ehmann, A. O. F. Jones, I. Salzmann, Q. Shen, C. Teichert, C. R. xe, G. Schweicher, Y. H. Geerts, R. Resel, M. Sferrazza and O. Werzer, *Soft Matter*, 2017, **13**, 2322-2329.
44. K. Kim, J. Chung, G. Hwang, O. Kwon and J. S. Lee, *ACS Nano*, 2011, **5**, 8700-8709.
45. H. S. Carslaw and J. C. Jaeger, *Conduction of Heat in Solids*, Oxford University Press, 1959.
46. P. L. Kapitza, *J. Phys. (Moscow)*, 1941, **4**, 181.
47. S. M. Lee and D. G. Cahill, *J. Appl. Phys.*, 1997, **81**, 2590-2595.
48. H.-K. Lyeo and D. G. Cahill, *Physical Review B*, 2006, **73**, 234.

49. Z. Chen, W. Jang, W. Bao, C. N. Lau and C. Dames, *Appl. Phys. Lett.*, 2009, **95**, 161910.
50. M. D. Losego, M. E. Grady, N. R. Sottos, D. G. Cahill and P. V. Braun, *Nature Materials*, 2012, **11**, 502-506.
51. F. Menges, H. Riel, A. Stemmer, C. Dimitrakopoulos and B. Gotsmann, *Phys. Rev. Lett.*, 2013, **111**, 205901.
52. R. Cheaito, J. T. Gaskins, M. E. Caplan, B. F. Donovan, B. M. Foley, A. Giri, J. C. Duda, C. J. Szwejkowski, C. Constantin, H. J. Brown-Shaklee, J. F. Ihlefeld and P. E. Hopkins, *Physical Review B*, 2015, **91**, 035432.
53. Z. Ding, J.-W. Jiang, Q.-X. Pei and Y.-W. Zhang, *Nanotechnology*, 2015, **26**, 065703.
54. C. Evangeli, J. Spiece, S. Sangtarash, A. J. Molina Mendoza, M. Mucientes, T. Mueller, C. Lambert, H. Sadeghi and O. Kolosov, *Adv. Electron. Mater.*, 2019, **7**, 1900331.
55. J. R. Dryden, *J. Heat Transfer*, 1983, **105**, 408-410.
56. D. Wang, W. Shi, J. Chen, J. Xi and Z. Shuai, *Physical Chemistry Chemical Physics*, 2012, **14**, 16505-16520.
57. G. T. Craven and A. Nitzan, *Nano Lett*, 2019, **20**, 989-993.
58. Y. He, I. Savić, D. Donadio and G. Galli, *Phys. Chem. Chem. Phys.*, 2012, **14**, 16209-16222.
59. D. P. Sellan, E. S. Landry, J. E. Turney, A. J. H. McGaughey and C. H. Amon, *Phys. Rev. B*, 2010, **81**, 214305.
60. G. Schweicher, G. D'Avino, M. T. Ruggiero, D. J. Harkin, K. Broch, D. Venkateshvaran, G. Liu, A. Richard, C. Ruzié, J. Armstrong, A. R. Kennedy, K. Shankland, K. Takimiya, Y. H. Geerts, J. A. Zeitler, S. Fratini and H. Sirringhaus, *Adv. Mater.*, 2019, **31**, 1902407.
61. H. Zhang, Y. Yao, M. M. Payne, J. E. Anthony and J. W. Brill, *Appl. Phys. Lett.*, 2014, **105**, 073302.

62. J. W. Brill, M. Shahi, M. M. Payne, J. Edberg, Y. Yao, X. Crispin and J. E. Anthony, *J. Appl. Phys.*, 2015, **118**, 235501.

63. Y. S. Muzychka, M. M. Yovanovich and J. R. Culham, *Journal of Thermophysics and Heat Transfer*, 2004, **18**, 45-51.

64. S. Haas, Y. Takahashi, K. Takimiya and T. Hasegawa, *Appl. Phys. Lett.*, 2009, **95**, 022111.

65. W. Xie, K. Willa, Y. Wu, R. Häusermann, K. Takimiya, B. Batlogg and C. D. Frisbie, *Adv. Mater.*, 2013, **25**, 3478-3484.

66. E. Selezneva, A. Vercouter and G. Schweicher, *et al., to be submitted*.

67. R. J. Bell, in Methods in Computational Physics: Advances in Research and Applications, ed. G. Gilat, Elsevier, 1976, vol. 15, pp. 215-276.

68. A. Cappai, A. Antidormi, A. Bosin, D. Narducci, L. Colombo and C. Melis, *Physical Review Materials*, 2020, **4**, 035401.

69. M. Saito, I. Osaka, E. Miyazaki, K. Takimiya, H. Kuwabara and M. Ikeda, *Tetrahedron Letters*, 2011, **52**, 285-288.

70. C. Grigoriadis, C. Niebel, C. Ruzié, Y. H. Geerts and G. Floudas, *J. Phys. Chem. B*, 2014, **118**, 1443-1451.

71. A. Majumdar, Annu. Rev. Mater. Sci., 1999, **29**, 505-585.

72. S. Gomès, A. Assy and P.-O. Chapuis, *phys. stat. sol. (a)*, 2015, **212**, 477-494.

73. S. Lefèvre and S. Volz, *Rev. Sci. Instrum.*, 2005, **76**, 033701.

74. S. Plimpton, Journal of Computational Physics, 1995, **117**, 1-19.

# Thermal conductivity of benzothieno-benzothiophene derivatives at the nanoscale

*Magatte N. Gueye,[a,#] Alexandre Vercouter,[b,#] Rémy Jouclas,[c] David Guérin,[a]*

*Vincent Lemaur,[b] Guillaume Schweicher[c,]*,[c] Stéphane Lenfant,[a]*

*Aleandro Antidormi,[d] Yves Geerts,[c,e] Claudio Melis,[f]*

*Jérôme Cornil[b,]* and Dominique Vuillaume[a,]* *

a) Institute for Electronics Microelectronics and Nanotechnology (IEMN), CNRS,

Av. Poincaré, Villeneuve d'Ascq, France.

b) Laboratory for Chemistry of Novel Materials, University of Mons,

Place du Parc 20, Mons, Belgium.

c) Laboratoire de Chimie des Polymères, Faculté des Sciences, Université Libre de Bruxelles (ULB),

Boulevard du Triomphe, 1050, Brussels, Belgium.

d) Catalan Institute of Nanoscience and Nanotechnology (ICN2), CSIC and BIST, Campus UAB,

Bellaterra, 08193, Barcelona, Spain.

e) International Solvay Institutes for Physics and Chemistry, Brussels, Belgium.

f) Dipartimento di Fisica, Universita di Cagliari, Cittadella Universitaria, 09042 Monserrato (Ca),

Italy.

Corresponding authors: dominique.vuillaume@iemn.fr ; guillaume.schweicher@ulb.ac.be  ;

jerome.cornil@umons.ac.be

# Supporting information

1. Sample fabrication
2. SThM measurements
3. Theoretical methods
4. Additional SThM images
5. NP-SThM calibration
6. Estimation of the constriction thermal resistance of the Si/SiO$_2$ substrate

7. Estimation of the radius of the thermal contact between tip and surface
8. Fitted parameters of Eq. 3.

**1. Sample fabrication**

The Si-n+(highly doped ~$10^{-3}$ Ω.cm)/$SiO_2$ substrate (200 - 500 nm of thermal $SiO_2$) were first cleaned in $CH_2Cl_2$ (3mn, sonicated), in IPA (isopropyl alcohol) solution (3mn, sonicated) and then submitted during 10 mn to UV-ozone cleaning. They were used immediately. The C8-BTBT-C8 and BTBT (scheme 1) solutions (5mg/mL in toluene for BTBT and 5 mg/mL in tetrahydrofuran for C8-BTBT-C8) were spin coated at 1000 rpm for 9s followed by 30 s at 1500 rpm, or directly at 4500 rpm for 30 s. We used two methods to deposit a drop of the solution onto the substrate: we dropped the solution on the substrate and started the spin-coater (method A) or we dropped the solution while the spin coater was already running (method B). Table S1 summarizes the spin-coating conditions for the 9 samples. The films are not uniform and we have indicated the maximum thickness of the films as measured from AFM images (see Figs. S1 and S2).

|    | C8-BTBT-C8 |        | BTBT |        |
|----|------------|--------|------|--------|
| #1 | 1000 rpm/9 s + 1500 rpm/30s (method A) | 410 nm | 4500 rpm/30s (method B) | 360 nm |
| #2 | 1000 rpm/9 s + 1500 rpm/30s (method B) | 160 nm | 4500 rpm/30s (method A) | 360 nm |
| #3 | 4500 rpm/30s (method A) | 190 nm | 4500 rpm/30s (method B) | 320 nm |
| #4 | 4500 rpm/30s (method B) | 90 nm | 4500 rpm/30s (method A) | 390 nm |
| #5 | 1000 rpm/9 s + 1500 rpm/30s (method B) | 390 nm | | |

**Table S1.** *Spin-coating parameters and maximum film thickness (rounded to the unit) estimated from AFM measurements.*

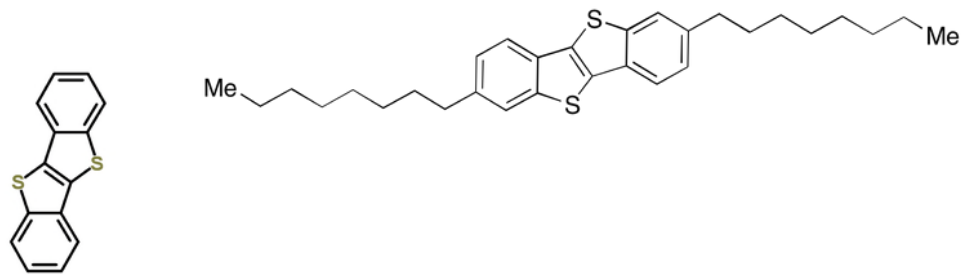

***Scheme S1.*** *Chemical structures: BTBT (left) and C8-BTBT-C8 (right), Me=CH$_3$.*

## 2. SThM measurements

***Topographic and thermal images***. In the scanning mode, the topography and thermal voltage were recorded simultaneously. We used both the DC method or the 3ω-SThM. The SThM tip is a lithographed Pd wire, i.e. resistance, inserted in a Wheatstone bridge, which is supplied by a DC voltage, V$_{DC}$. The tip is used to heat the sample (the higher the DC voltage, the higher the heat flux and the tip temperature) and to measure the tip/surface temperature, in the so-called active mode. The tip resistance is measured by the Wheatstone bridge connected to a voltage amplifier. From this output SThM voltage, we deduced the tip resistance, which is calibrated versus temperature. Thus, the SThM voltage recorded at the amplifier output can be converted to tip/sample temperature. To minimize parasitic contributions (air conduction, radiation) and thus increase the sensitivity of the tip/sample contribution, we also use the 3ω-SThM method,[1, 2] in which the Wheatstone bridge is supplied by an AC voltage (0.2 V at 1 kHz) and we measure the third harmonic signal at the Wheatstone bridge output with a lock-in.

***Null-point SThM method.*** The null-point SThM[3] was used at selected points on the organic films (organic domains) and/or the uncovered Si/SiO$_2$ substrate. From a previously recorded topographic/thermal image a zone of interest is selected where we define a 5x5 grid, each point spaced by 10 nm. At each point of the grid, in the z-trace mode (approach and retract) we recorded the SThM voltage versus distance curve ($V_{SThM}$-z). At the transition from a non-contact (NC, tip very near the surface) to a contact (C, tip on the surface) situation, we observe a temperature jump, $T_{NC} - T_C$, which is used to determine the sample thermal conductivity according to the protocol described in Ref. 3. The temperature jump is measured from the approach trace only (to avoid any artifact due to well-known adhesion hysteresis of the retract curve) and averaged over the 25 recorded $V_{SThM}$-z traces. This differential method is suitable to remove the parasitic contributions (air conduction, etc...): at the contact (C) both the sample and parasitic thermal contributions govern the $V_{SThM}$ signal, whereas, just before physical tip contact (NC), only the parasitic thermal contributions are involved. The plot of the temperature jump, $T_{NC} - T_C$, versus the sample temperature at contact $T_C$ is linear and its slope is inversely proportional to the thermal conductivity. The tip-sample temperature $T_C$ increases with the supply voltage of the Wheatstone bridge $V_{DC}$ (typically from 0.4 to 1 V).

***Thermal flux.*** The determination of the thermal resistance from the ratio of the thermal voltage measured on the organic domain versus over the substrate is valid under the hypothesis that all the electrical power delivered to the tip is dissipated into the sample and the thermal flux is constant.[4, 5] Albeit not strictly verified, this is justified, in first approximation, since the constriction resistances on the two zones are quite similar around $10^7$ K W$^{-1}$ ($R_{sub}=1/4r\kappa_{OX}$ = 9x10$^6$ K W$^{-1}$, $R_{org}=1/4r\kappa_{org}$ = 2-5x10$^7$ K W$^{-1}$, considering the values of $\kappa_{org}$ determined by the NP-SThM). However, in addition to taking into account the interface resistances (see text), a more precise analysis of data in Fig. 3 implies that a correction factor should be applied to determine the ratio $R^*_{org}/R_{sub}$ from the ratio of the SThM voltages, $R^*_{org}(t)/R_{sub} = \left(V_{SThm-org}/V_{SThM-sub}\right)\left(\dot{Q}_{sub}/\dot{Q}_{org}\right)$. Since the thermal conductivity is lower for C8-BTBT-C8 than for BTBT, the correction factor $\dot{Q}_{sub}/\dot{Q}_{org}$ should be higher for C8-BTBT-C8.

## 3. Theoretical methods

***Molecular Dynamics.*** A wide range of Molecular Dynamics (MD) simulations has been developed to assess the lattice thermal conductivity of a plethora of materials and complex nanostructures.[6, 7] Among them, the Green-Kubo (GK) formalism[8] based on the fluctuation-dissipation theorem provides a direct access to the (off-)diagonal elements of the two-dimensional thermal conductivity tensor. Nevertheless, ensuring the proper convergence of the auto- and cross-correlation functions of the heat current may be a non-trivial task, especially for soft molecular systems. A second class of techniques built on a Non-Equilibrium Molecular Dynamics (NEMD) scheme[9] requires important computational resources to artificially maintain the system out of equilibrium while estimating its thermal conductivity in a stationary state. The key steps of the AEMD methodology are the following: (i) applying a perfectly monitored thermal pulse on a simulation box; (ii) recovering the initial thermodynamic equilibrium during a fast transient regime; (iii) fitting the time-decaying temperature difference between the right and left parts of the system from a reliable solution of the one-dimensional heat equation $\frac{\partial T}{\partial t} = D \frac{\partial^2 T}{\partial x^2}$ in order to evaluate the thermal diffusivity, $D$.[10] This alternative scheme has the benefit to be far less time-consuming than the previously cited methods due to the rapid dissipation of the thermal gradient. The AEMD approach also contrasts with both the Equilibrium Molecular Dynamics (EMD) and the Non Equilibrium Molecular Dynamics (NEMD) techniques by estimating the phononic contribution to the thermal transport with no need for any instantaneous heat flux; moreover, the definition of the heat flux in the LAMMPS software[11] has been recently identified as physically inconsistent for the study of molecular systems.[12, 13] With the AEMD approach, the thermal conductivity $\kappa = D\rho C_P$ is finally obtained, provided the density, $\rho$, and the specific heat, $C_P$, of the system are known.[10] The parameter $C_P$ is typically estimated to be 3*R* (with *R* the ideal gas constant), as predicted by the Dulong-Petit model.[14] Note that no quantum corrections are taken into account because MD simulations are performed at 300K while many OSCs are characterized by a very low Debye temperature $\theta_D$ [15, 16] so that the classical approximation used in the Dulong-Petit relationship is still valid. It is also of prime importance to account for the size-dependence of the lattice thermal conductivity deduced from this approach since phonons having a mean free path larger than the cell dimension do not effectively contribute to Κ. An extrapolation procedure is thus needed to get rid of these size effects[17] and to extract a quasi-length-free lattice thermal conductivity from the linear regression of 1/Κ versus 1/L (L the length of the box size along the direction of heat propagation). We have carefully reparameterized the torsional potential terms against sophisticated quantum-chemical calculations, with atomic charges calculated at the 1.14*CM1A-LBCC level.[18] The implementation of the force field was insured by the LigParGen free server.[19] Supercells ranging from 10 to 52 units cell have been generated along each direction of interest for heat transport and periodic boundary conditions were applied along the main three axes. First, atomic positions were optimized at 0 K while keeping fixed the cell parameters before relaxing the whole simulation box during a second energy minimization. Afterwards, systems have been aged in the canonical NVT ensemble (mole (N), volume (V) and temperature (T)) and in the isothermal–isobaric NPT ensemble (mole (N), pressure (P) and temperature (T)) during 1 ns in ambient conditions, using a Nose-Hoover thermostat and barostat. Then, a step-like temperature profile was created by simultaneously "freezing" atomic motions in one half of the simulation box while heating up [cooling down, respectively] the second part of the system alternatively at < T1 > = 362.5 K (< T2 > = 237.5 K, respectively) via NVT simulations (∼ 250 ps). Finally, an NVE (mole (N), volume (V) and energy (E)) simulation lasting up to ∼ 3 ns was achieved to dissipate the initial 125 K thermal gradient. Throughout a post-processing step, the monitored temperature offset $\Delta T = \langle T_1 \rangle - \langle T_2 \rangle$ was fitted by a sum of exponentials expressed as $\sum_{n=1}^{5} C_n e^{-\alpha_n^2 D t}$, where $C_n$ and $\alpha_n$ are coefficients depending on initial conditions. A fitting function based on five terms is the best compromise between reasonable computational time and sufficient accuracy for the estimation of the lattice thermal conductivity.

***Estimation of the participation ratio.*** In order to characterize the vibrational properties of both BTBT and C8-BTBT-C8, we performed the diagonalization of the corresponding dynamical matrix given by:

$$D_{i\alpha,j\beta} = \frac{-1}{\sqrt{m_i m_j}} \frac{\partial F_{i\alpha}}{\partial r_{j\beta}} \qquad (s1)$$

where $m_i$ is the mass of the i[th] atom, $F_{i\alpha}$ is the force on the i[th] atom along the α direction due to a displacement of atom j along the β direction. Latin indices are used for labelling atoms while Greek letters indicate the (x,y,z) Cartesian components. The diagonalization of the dynamical matrix was obtained by using the SLEPc library[20] in order to obtain the eigenvectors $e_s$ and eigenvalues $\omega^2_s$ where s = 1, . . . ,3N counts the system eigenmodes. The calculation of the force first derivatives have been performed using a numerical finite difference procedure by considering an atomic displacement as small as 5 x 10[-4] Å. Once the eigenvectors have been calculated, we estimated the participation ratio (PR) as:[21]

$$PR = \frac{1}{N} \frac{\left(\sum_{i=1}^{N} e_{i,s}^2\right)^2}{\sum_{i=1}^{N} e_{i,s}^4} \qquad (s2)$$

PR yields an estimation of the contribution of a subgroup of atoms in a specific vibrational mode. The actual localized or extended character of that mode is related to the actual PR value: PR ~ 1 in the case of extended modes while PR has a smaller value for localized modes. It is worth noticing that, according to its very definition, a PR exactly equal to unity is obtained solely for vibrational modes in ideally perfect crystalline systems, in which the atomic displacements are perfectly periodic within the sample. In general, the PR values for extended modes in non-crystalline systems lie in the 0.4- 0.6 interval.

## 4. Additional SThM images

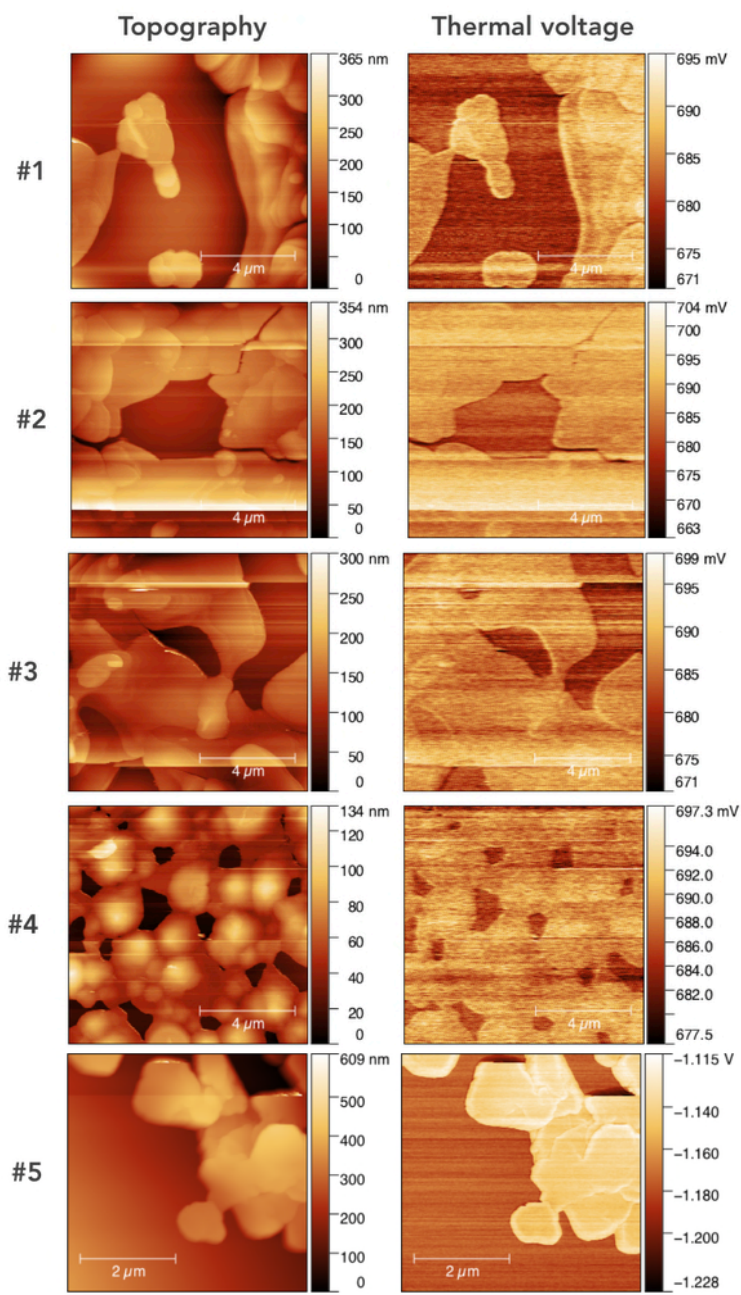

***Figures S1.*** *Typical topographic and thermal voltage images of the five C8-BTBT-C8 measured by 3ω-SThM ($V_{AC}$ = 0.2 V, 1 kHz), samples #1-4, and by DC-SThM ($V_{DC}$ = 0.5V), sample #5.*

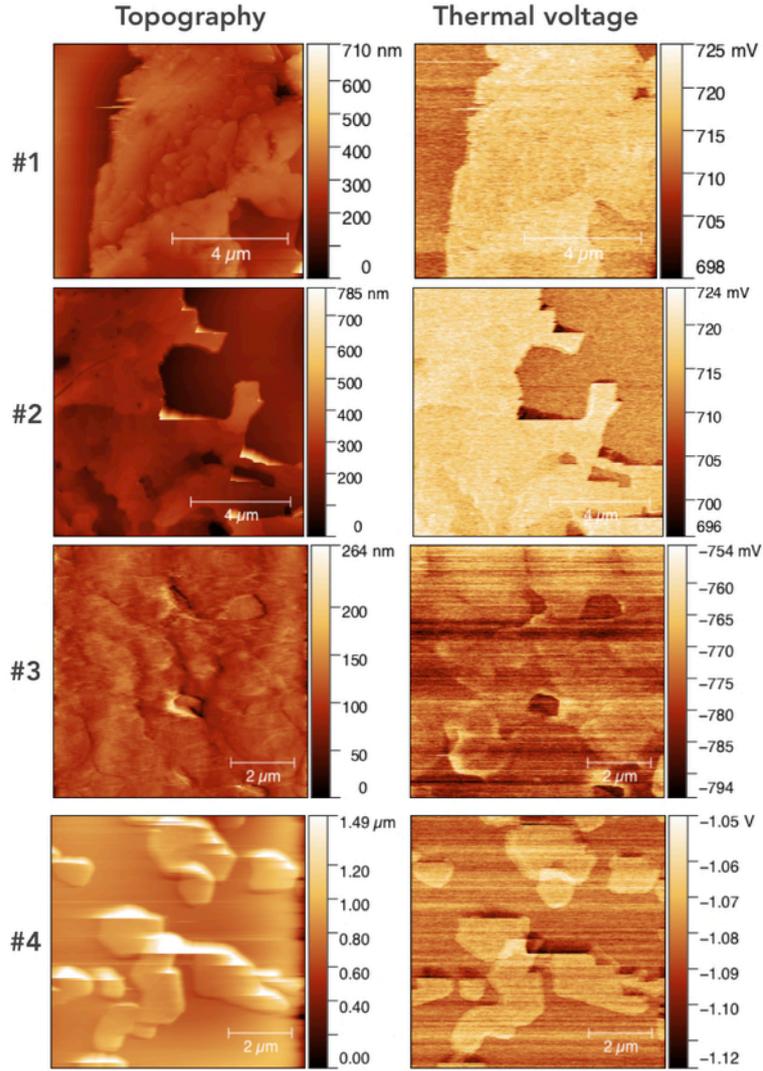

*Figure S2.* Typical topographic and thermal voltage images of the BTBT samples measured by 3ω-SThM ($V_{AC}$ = 0.2 V, 1 kHz) for samples #1 and 2 and samples #3,4 by DC-SThM ($V_{DC}$ = 0.6V).

## 5. NP-SThM calibration.

The sample temperature is calculated from the measured thermal voltage by

$$V_{SThM} = GR_{tip}\left(T_{sample}\right) = G\left[R_{amb} + \lambda\left(T_{sample} - T_{amb}\right)\right] \qquad \text{(Eq.S1)}$$

where G is the known transfer function of the Wheatstone bridge and the amplifier gain (1.08 x $V_{DC}$ V/Ω) for the DC-SThM methods,[22] 8.2x10$^{-2}$ x $V_{AC}$ V/Ω for the 3ω-SThM. The tip resistance $R_{tip}$ varies linearly with sample temperature, and has been calibrated given $R_{amb}$= 290.2 Ω and λ = 0.27 Ω/°C.

To determine the thermal conductivity from data in Fig. 3 and using Eq. (2), we calibrated the null-point SThM according to the protocol in Ref. 3. The same $T_C$ vs. $T_{NC}$-$T_C$ measurements were done on two materials with well-known thermal conductivity (below) : a glass slide (1.3 W/m.K) and a low doped silicon wafer with its native oxide (150 W/m.K). From a linear fit on the data, we get α = 25.6 W m$^{-1}$ K$^{-1}$ and β = 21.6 K/K.

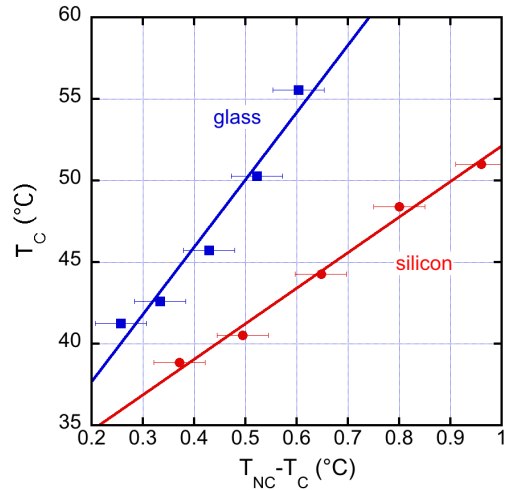

**Figure S3.** $T_C$ vs. $T_{NC}$ - $T_C$ plot for the two reference materials. Solid lines are the linear fits.

## 6. Estimation of the constriction thermal resistance of the Si/SiO$_2$ substrate

The effective thermal conductivity measured by NP-SThM is the one of the bulk SiO$_2$ (see Fig. 6-b in Ref. 3 and below) if the SiO$_2$ thickness is larger than ∼ 100 nm (here 200-500 nm).[3] This behavior is in agreement with a simple analytical model derived by Dryden[23] for film with *t/r* > 2 (*r* being the radius of the SThM tip thermal contact ($r \approx 20$ nm, see below). Thus, we used for the Si/SiO$_2$ substrate $R_{sub}=1/4r\mathcal{K}_{OX}$ = 9x10$^6$ K W$^{-1}$ with $\mathcal{K}_{OX}$ the "bulk" SiO$_2$ thermal conductivity (1.4 W m$^{-1}$ K$^{-1}$).

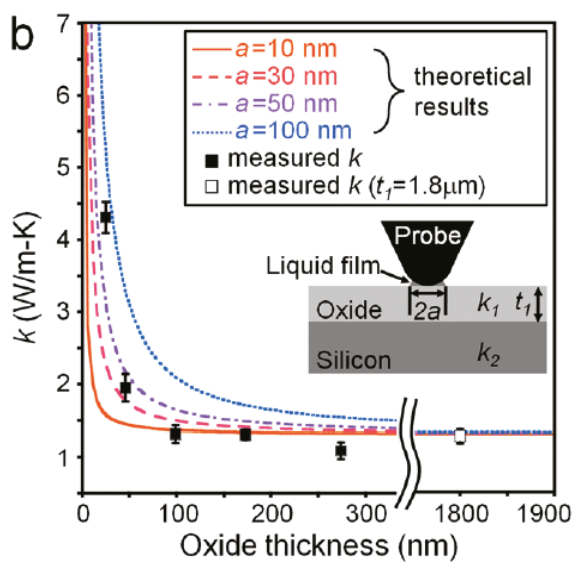

***Figure S4.*** *Variation of the measured thermal conductivity of a silicon substrate covered by a film of SiO$_2$ with various thickness (Reprinted with permission from Kim, K.; Chung, J.; Hwang, G.; Kwon, O.; Lee, J. S. ACS Nano 2011, 5, 8700–8709. Copyright (2011) American Chemical Society).*

**7. Estimation of the radius of the thermal contact between tip and surface.**

The thermal contact radius is calculated following the approach reported in Ref. 24 taken into account the mechanical tip radius $r_{tip}$ and the size of the water meniscus at the tip/surface interface. The thermal radius of the thermal contact is given by [25]

$$r_{th} = 2.08 \sqrt{\frac{-r_{tip} \cos\theta}{\ln\varphi}} \qquad \text{(Eq. S2)}$$

with $r_{tip}$ = 100 nm (data from Bruker), the relative humidity φ = 0.35-0.4 (air-conditioned laboratory, values checked during the measurements) and the contact angle of the concave meniscus between the tip and the surface θ ≈ 30° as measured for π-conjugated molecular crystals in Ref. 26. We get r ≈ 20 nm. The water meniscus contact angle depends on the surface energy of the sample, and thus should, in principle, not be the same on the organic domains and the uncovered $SiO_2$ surface, the later being more hydrophilic than the organic materials. However, we cannot perform standard water contact angle measurements inside the micrometer size uncovered $SiO_2$ (see Fig. S1 and S2), and we consider the same value of 30° in both cases. Moreover, it is likely that the $SiO_2$ surface is covered by organic contaminants from the spin-coating deposition technique.

**8. Fitted parameters of Eq. 3.**

Table S2 gives the values of the fitted parameters ($\kappa_{org}$) for the data shown in Fig. 4 (main text).

|  | C8-BTBT-C8 | BTBT |
|---|---|---|
|  | $\kappa_{org}$ (W $m^{-1}$ $K^{-1}$) | $\kappa_{org}$ (W $m^{-1}$ $K^{-1}$) |
| sample #1 | 1.348 | 1.371 |
| sample #2 | 1.352 | 1.373 |
| sample #3 | 1.359 | 1.366 |
| sample #4 | 1.355 | 1.372 |

*Table S2. Fit parameters (Fig. 4, Eq. 3)*

We get the average values $\kappa_{BTBT}$ = 1.37 ± 0.01 W $m^{-1}$ $K^{-1}$ and $\kappa_{C8-BTBT-C8}$ = 1.35 ± 0.01 W $m^{-1}$ $K^{-1}$. We note that a values of r = 80 nm has been used to these fits. Using r = 20 nm gave poor fits. The value r = 80 nm should indicate a more hydrophilic surface (Eq. S2, lower θ, but Eq. (S2) has a limit r = 21.7 nm for θ=0°, thus this simple model is inappropriate in the case of highly hydrophilic surface). Nevertheless, this uncertainty in the determination of *r* does not change the conclusion of the NP-SThM measurements that the thermal conductance of BTBT is higher than for C8-BTBT-C8 (r is included in the calibration parameters).

**References**

1.	Lefèvre, S.; Volz, S., 3ω-scanning thermal microscope. *Rev. Sci. Instrum.* **2005,** *76* (3), 033701.

2.	Bodzenta, J.; Juszczyk, J.; Chirtoc, M., Quantitative scanning thermal microscopy based on determination of thermal probe dynamic resistance. *Rev. Sci. Instrum.* **2013,** *84* (9), 093702.

3.	Kim, K.; Chung, J.; Hwang, G.; Kwon, O.; Lee, J. S., Quantitative measurement with scanning thermal microscope by preventing the distortion due to the heat transfer through the air. *ACS Nano* **2011,** *5* (11), 8700-8709.

4.	Menges, F.; Riel, H.; Stemmer, A.; Gotsmann, B., Quantitative Thermometry of Nanoscale Hot Spots. *Nano Lett* **2012,** *12*, 596-601.

5.	Meier, T.; Menges, F.; Nirmalraj, P.; Hölscher, H.; Riel, H.; Gotsmann, B., Length-Dependent Thermal Transport along Molecular Chains. *Phys. Rev. Lett.* **2014,** *113* (6), 060801.

6.	Fugallo, G.; Colombo, L., Calculating lattice thermal conductivity: a synopsis. *Physica Scripta* **2018,** *93* (4), 043002.

7.	Puligheddu, M.; Xia, Y.; Chan, M.; Galli, G., Computational prediction of lattice thermal conductivity: A comparison of molecular dynamics and Boltzmann transport approaches. *Physical Review Materials* **2019,** *3* (8), 085401.

8.	Kubo, R.; Toda, M.; Hashitsume, N., *Statistical Physics II*. Spinger: Berlin, 1985.

9.	Müller-Plathe, F., A simple nonequilibrium molecular dynamics method for calculating the thermal conductivity. *The Journal of Chemical Physics* **1997,** *106* (14), 6082-6085.

10.	He, Y.; Savić, I.; Donadio, D.; Galli, G., Lattice thermal conductivity of semiconducting bulk materials: atomistic simulations. *Physical Chemistry Chemical Physics* **2012,** *14* (47), 16209-16222.

11. Plimpton, S., Fast Parallel Algorithms for Short-Range Molecular Dynamics. *Journal of Computational Physics* **1995,** *117* (1), 1-19.

12. Boone, P.; Babaei, H.; Wilmer, C. E., Heat Flux for Many-Body Interactions: Corrections to LAMMPS. *Journal of Chemical Theory and Computation* **2019,** *15* (10), 5579-5587.

13. Surblys, D.; Matsubara, H.; Kikugawa, G.; Ohara, T., Application of atomic stress to compute heat flux via molecular dynamics for systems with many-body interactions. *Physical Review E* **2019,** *99* (5), 051301.

14. Ashcroft, N. W.; Mermin, N. D., *Solid State Physics*. HoltSander: London, 1976.

15. Ramadoss, P.; Buvaneswari, N., Thermal Properties of Some Organic Liquids Using Ultrasonic Velocity Measurements. *E-Journal of Chemistry* **2011,** *8*, 120260.

16. Babavali, S. F.; Nori, T. S.; Srinivasu, C., Acoustical effective Debye temperature studies in heterocyclic aromatic functional materials of organic liquid mixtures at four different temperatures T = (303.15, 308.15, 313.15 and 318.15) K. *Materials Today: Proceedings* **2019,** *18*, 2073-2076.

17. Sellan, D. P.; Landry, E. S.; Turney, J. E.; McGaughey, A. J. H.; Amon, C. H., Size effects in molecular dynamics thermal conductivity predictions. *Physical Review B* **2010,** *81* (21), 214305.

18. Dodda, L. S.; Vilseck, J. Z.; Tirado-Rives, J.; Jorgensen, W. L., 1.14*CM1A-LBCC: Localized Bond-Charge Corrected CM1A Charges for Condensed-Phase Simulations. *The Journal of Physical Chemistry B* **2017,** *121* (15), 3864-3870.

19. Dodda, L. S.; Cabeza de Vaca, I.; Tirado-Rives, J.; Jorgensen, W. L., LigParGen web server: an automatic OPLS-AA parameter generator for organic ligands. *Nucleic Acids Research* **2017,** *45* (W1), W331-W336.

20. Hernandez, V.; Roman, J. E.; Vidal, V., SLEPc: A Scalable and Flexible Toolkit for the Solution of Eigenvalue Problem. *ACM Trans. Math. Softw.* **2005,** *31*, 351-362.

21. Antidormi, A.; Cartoixà, X.; Colombo, L., Nature of microscopic heat carriers in nanoporous silicon. *Physical Review Materials* **2018,** *2* (5), 056001.

22. Anasys-Instruments SThM installation and operation manual; 2008.

23. Dryden, J. R., The Effect of a Surface Coating on the Constriction Resistance of a Spot on an Infinite Half-Plane. *J. Heat Transfer* **1983,** *105*, 408-410.

24. Luo, K.; Shi, Z.; Varesi, J.; Majumdar, A., Sensor nanofabrication, performance, and conduction mechanisms in scanning thermal microscopy. *Journal of Vacuum Science & Technology B: Microelectronics and Nanometer Structures* **1997,** *15* (2), 349-360.

25. Israelachvili, J. N., Intermolecular and surface forces; 3rd ed. Academic Press: Amsterdam, 2011.

26. Zhang, Y.; Zhang, C.; Wei, D.; Bai, X.; Xu, X., Nanoscale thermal mapping of few-layer organic crystal. *CrystEngComm* **2019,** *21*, 5402-5409.